\documentclass[]{spie}  

\usepackage{amsmath,amsfonts,amssymb}
\usepackage{graphicx}
\usepackage[colorlinks=true, allcolors=blue]{hyperref}
\usepackage{astro_bib_macro}

\usepackage{booktabs}

\usepackage{listings}

\usepackage{enumitem}
\newlist{arrowlist}{itemize}{1}
\setlist[arrowlist]{label=$\Rightarrow$}

\title{Prototyping liquid-crystal coronagraphs for exo-Earth imaging}

\author{Iva Laginja\supit{a,b}, David Doelman\supit{a,c}, Frans Snik\supit{a}, Pierre Baudoz\supit{b}, Felix Bettonvil\supit{a}, Jeroen H. H. Rietjens\supit{c}, Chris N. van Dijk\supit{d}, Kristien Peeters\supit{d}, Alexander Eigenraam\supit{c}, Erin Pougheon\supit{b}, Tom van der Wielen\supit{a}, Marco Esposito\supit{d}, Thomas Wijnen\supit{a}, Mariya Krasteva\supit{e}, Matteo Taccola\supit{e}
\skiplinehalf
\supit{a} NOVA/Leiden University, Einsteinweg 55, 2333 CC Leiden, The Netherlands\\
\supit{b} LESIA, Observatoire de Paris, Université PSL, CNRS, Sorbonne Université, Université Paris
Cité, 5 place Jules Janssen, 92195 Meudon, France\\
\supit{c} SRON Netherlands Institute for Space Research, Niels Bohrweg 4, 2333 CA Leiden, The Netherlands\\
\supit{d} cosine Remote Sensing B.V., Warmonderweg 14, 2171 AH Sassenheim, The Netherlands\\
\supit{e} European Space Agency (ESA), European Space Research and
Technology Centre (ESTEC), Keplerlaan 1, 2201 AZ Noordwijk,
The Netherlands\\
}

\authorinfo{Further author information, send correspondence to Iva Laginja: E-mail: iva.laginja@obspm.fr}

\begin{document} 
\maketitle

\begin{abstract}
This paper presents initial results from the ESA-funded ``SUPPPPRESS'' project, which aims to develop high-performance liquid-crystal coronagraphs for direct imaging of Earth-like exoplanets in reflected light. The project focuses on addressing the significant challenge of polarization leakage in vector vortex coronagraphs (VVCs). We utilize newly manufactured multi-grating, liquid-crystal VVCs, consisting of a two- or three-element stack of vortex and grating patterns, to reduce this leakage to the $10^{-10}$ contrast level. We detail the experimental setups, including calibration techniques with polarization microscopes and Mueller matrix ellipsometers to enhance the direct-write accuracy of the liquid-crystal patterns. The performance testing of these coronagraph masks will be conducted on the THD2 high-contrast imaging testbed in Paris.
\end{abstract}

\keywords{Vector vortex coronagraphs, high-contrast imaging, exo-Earth imaging, polarization, liquid-crystal patterns, diffraction gratings}

\section{Introduction}
\label{sec:introduction}

The search for exo-Earths, or Earth-like planets in the habitable zones of distant stars, is one of the most ambitious and challenging objectives in modern astronomy. Direct imaging and spectral characterization of these exoplanets require advanced instruments capable of high-contrast imaging (HCI) to differentiate the faint light from the planets against the overwhelming brightness of their host stars. Vortex coronagraphs, particularly vector vortex coronagraphs (VVCs)\cite{Mawet2009OpticalVectorial}, are crucial in this endeavor due to their ability to suppress starlight while allowing planetary light to pass through.

We report on the first results of the ESA-funded SUPPPPRESS project, which aims to develop liquid-crystal coronagraphs to directly image Earth-like exoplanets in reflected light. A high-performance liquid-crystal coronagraph\cite{Mawet2009OpticalVectorial} takes advantage of the unmatched wavefront error stability of space telescopes, to reach optimum contrast (1e-10 in reflected light) at high spatial resolution (a few mas) and sufficient sensitivity. A bandwidth of 20\% will allow the spectral characterization of such targets, and polarimetry in particular allows a direct measurement of cloud properties and liquid water through ocean glint\cite{vaughan2023chasing}.

We are preparing to create multi-grating VVCs, developed by Doelman et al. 2020\cite{doelman2020minimizing}, with two or three elements with grating patterns to reduce polarization leakage and compare them to standard VVCs. In particular, we investigate the direct-write accuracy of these new masks, and improve its accuracy with careful calibration using polarization microscopes and mueller matrix ellipsometers. With the HCI testing facility of choice being the in-air high-contrast “très haute dynamique” (THD2) testbed in Paris\cite{Baudoz2018OptimizationPerformanceMultideformable}, we describe the software and hardware upgrades performed for the SUPPPPRESS project.

\subsection{Science landscape}
\label{subsec:science}

The landscape of exoplanet science has been significantly shaped by the development and deployment of HCI instruments. These instruments aim to achieve the contrast levels required to detect and characterize exoplanets, which often have contrast ratios as extreme as $10^{-10}$ compared to their host stars. A prominent example of a future mission dedicated to this goal is the Habitable Worlds Observatory (HWO), a proposed space telescope designed to directly image Earth-like exoplanets in the habitable zones of nearby stars\cite{astro2020}.

The HWO represents the next generation of space telescopes, following in the footsteps of missions like the James Webb Space Telescope (JWST) and the Roman Space Telescope (RST). The HWO aims to provide unprecedented wavefront stability and contrast performance, enabling the direct detection of exoplanets\cite{Coyle2024}. To achieve this, HWO will incorporate advanced coronagraphs, including vortex coronagraphs, which are essential for suppressing the bright starlight and revealing the faint planetary signals.

The trade space for high-performing coronagraphs is vast\cite{Belikov2024}, with different designs offering varying levels of performance in terms of inner working angle (IWA), throughput, and spectral bandwidth. The IWA of a coronagraph is a critical parameter, as it defines the smallest angular separation at which the instrument can effectively suppress starlight and detect a planet. For exo-Earth imaging, achieving an IWA close to the diffraction limit of the telescope is crucial.

\begin{figure}
    \centering
   \includegraphics[width = 0.7\textwidth]{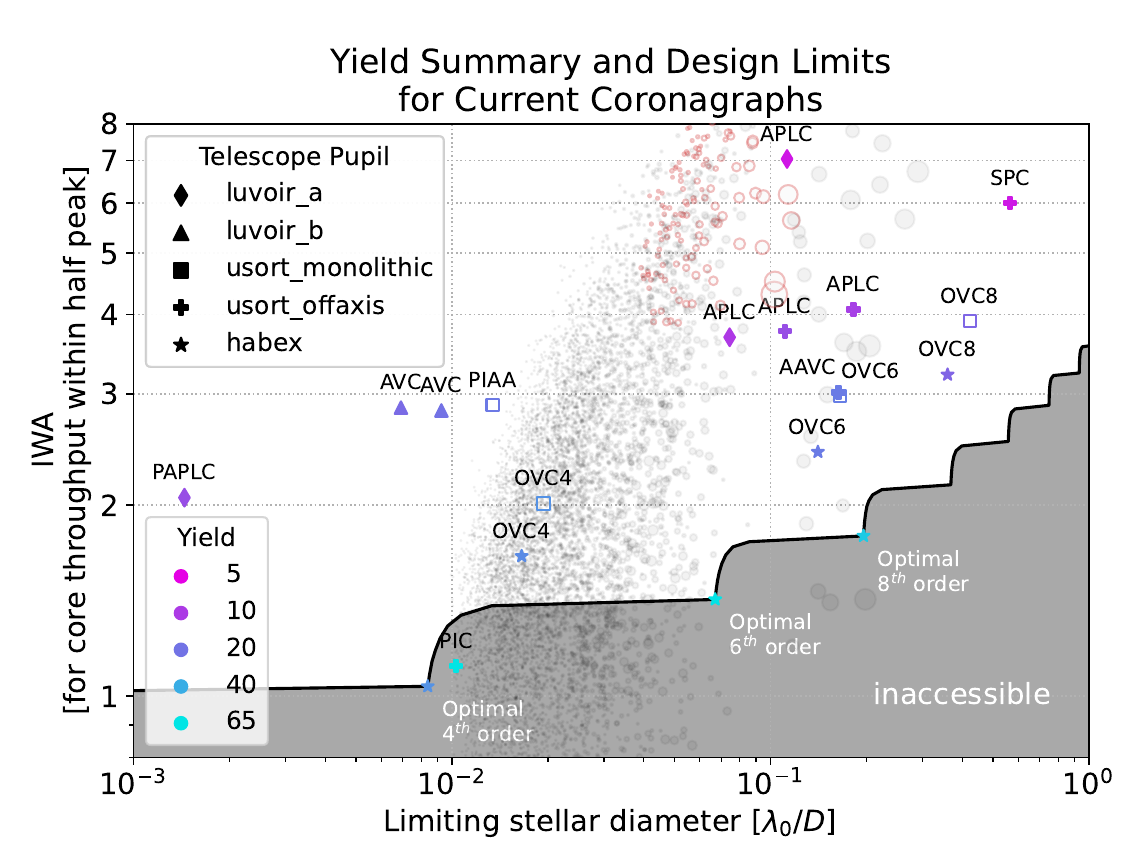}
   \caption[Figure] 
   {\label{fig:cds_theoretical_limits} 
   The trade space for high-performing coronagraphs adequate for exo-Earth imaging is large. Pictured here are the theoretical performance levels of various coronagraphs compared to theoretical limits in the gray area. Optical vortex coronagraphs (OVC) perform very close to the theoretical limit. Credit: Belikov et al. (2024)\cite{Belikov2024}.}
   \end{figure}

Optical vortex coronagraphs (OVCs) are particularly promising in this regard\cite{Foo2005}. Figure~\ref{fig:cds_theoretical_limits} illustrates the theoretical performance levels of various coronagraphs compared to the theoretical limits in the shaded grey area. OVCs perform very close to these limits, making them ideal candidates for HCI. These coronagraphs use a phase ramp to diffract starlight out of the beam path, allowing only the light from off-axis sources, such as exoplanets, to pass through the optical system and reach the detector.

\subsection{Liquid-crystal vector vortex coronagraphs}
\label{subsec:lc-VVCs}

\begin{figure}
    \centering
   \includegraphics[width = 0.8\textwidth]{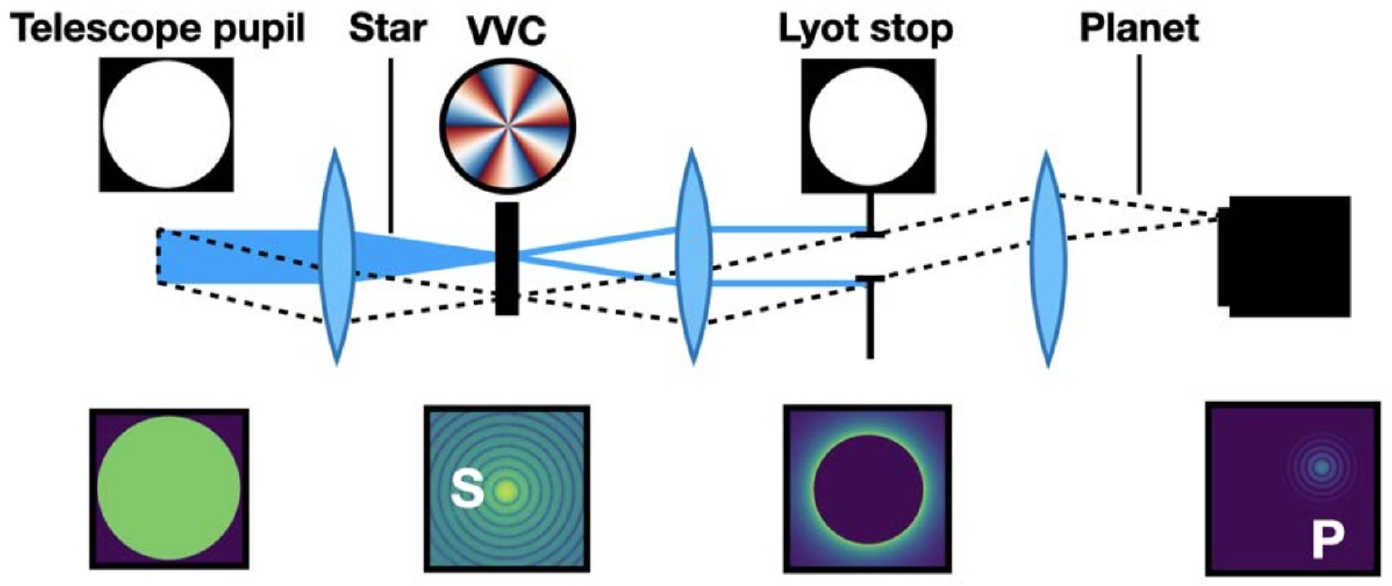}
   \caption[Figure] 
   {\label{fig:full_vvc} 
    Schematic of a perfect vector vortex coronagraph (VVC). On-axis stellar light (S) is diffracted by the VVC mask to outside the Lyot stop and suppressed by the Lyot stop. Off-axis planet light (P) is mostly unchanged by the VVC mask and can pass the Lyot stop to the science camera. The focal planes are shown in logarithmic scaling. Adapted from Doelman et al. (2020)\cite{doelman2020minimizing}.}
   \end{figure}

The VVC is a focal-plane coronagraph that uses a phase mask to diffract starlight out of the beam path, allowing the light from off-axis sources, such as exoplanets, to pass through unaltered\cite{Foo2005}. The key component of a VVC is the phase mask, which is designed with an azimuthal phase ramp of $2 \pi n$, where n is an integer known as the charge of the coronagraph. Higher charge values provide greater resilience against optical aberrations but also increase the IWA of the coronagraph. A layout of the classic VVC is displayed in Fig.~\ref{fig:full_vvc}.

Liquid-crystal materials are ideally suited for creating the complex phase patterns required for VVCs. These materials can be precisely controlled to produce the desired phase retardation, making them an excellent choice for manufacturing VVC masks. The geometric phase effect, or Pancharatnam-Berry phase\cite{Pancharatnam1955, Berry1984}, is exploited in these materials to achieve the necessary phase modulation.

Producing such achromatic phase masks is one of the primary advantages of using liquid-crystal materials in VVCs, as the phase imparted by the liquid-crystal pattern depends only on the orientation of the fast axis of the retarder and not on the wavelength of light. Liquid-crystal-based VVCs can maintain high performance across a broad spectral range up to 50\% bandwidth \cite{doelman2020minimizing}, making them highly suitable for broadband observations. This is particularly important for missions like HWO, which aim to characterize exoplanets across multiple wavelengths to gather comprehensive data on their atmospheres and potential habitability.

Despite their advantages, liquid-crystal VVCs are not without challenges. One of the main issues is polarization leakage, which occurs when the retardance of the mask deviates from the ideal half-wave value. This leakage results in a portion of the light not acquiring the intended phase shift and contaminating the coronagraphic image, thereby reducing the contrast and effectiveness of the coronagraph.

In the following section, we first recall the standard way of reducing this leakage by the use of polarization filters. Then, we introduce the alternative method of using polarization gratings to diffract this leakage off-axis. The development, manufacturing and testing of coronagraphs that integrate the latter solution is the goal of the SUPPPPRESS project.

\subsection{Polarization leakage in a vector vortex coronagraph}
\label{subsec:leakage}

Polarization leakage is a critical challenge in the implementation of vector vortex coronagraphs. It arises when the retardance of the liquid-crystal mask deviates from the ideal half-wave value of 180 degrees, causing a fraction of the incident light to not acquire the intended phase shift. This unmodulated light creates a regular, non-coronagraphic, on-axis point-spread function (PSF) that significantly reduces the contrast achievable by the coronagraph. The intensity polarization leakage fraction is $c_V^2$ (see Doelman et al., 2020\cite{doelman2020minimizing}), where $c_V$ depends on the deviation from half-wave, $\Delta \phi$, as given by
\begin{equation}
    c_V = \sin{\frac{\Delta \phi}{2}}.
\end{equation}

The primary cause of polarization leakage is the imperfect manufacturing of liquid-crystal masks. Achieving precise half-wave retardance across a significant wavelength range ($>10\%$) is difficult\cite{ruane2022broadband}. In this project we use self-aligning birefringent liquid crystals to tune the retardance of multi-layer liquid-crystal films\cite{Komanduri2013,Escuti2016}. By doping the liquid-crystal solutions we can modify the twist and with spin-coating we can change the thickness of individual layers to create multi-twist retarder (MTR) films. Generally, we use a Pancharatnam design where we use two layers with opposite twist and possibly a third layer in between without twist to minimize deviations from half-wave.

To second order, there are also spatially variant retardance deviations in the mask due to variations in thickness and alignment of the liquid-crystal layers. Even slight deviations from the ideal retardance can result in significant leakage. Other contributing factors include non-uniformities in the liquid-crystal material and environmental conditions affecting the birefringence.

\begin{figure}
    \centering
   \includegraphics[width = 0.6\textwidth]{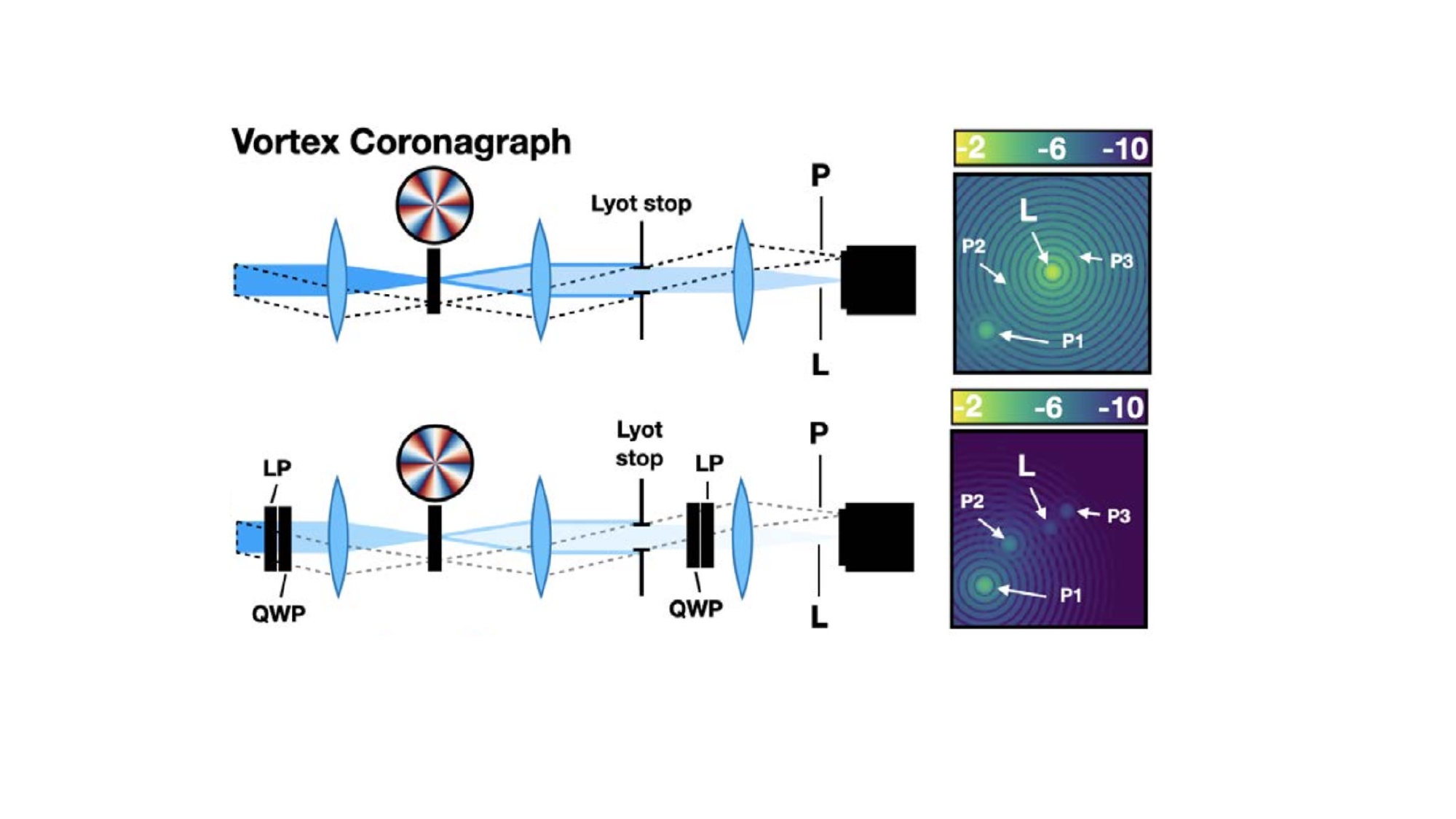}
   \caption[Figure] 
   {\label{fig:vvc_filtered_and_not} 
    Comparison of a non-perfect VVC with polarization leakage (L) with and without filtering. The focal-plane images are on logarithmic scale. We show three planets, where P1 has a contrast of $1\times10^{-5}$, P2: $1\times10^{-7}$, and P3: $1\times10^{-9}$. The two closest planets (P2,3) are mostly hidden by the polarization leakage for the VVC. Adapted from Doelman et al. (2023)\cite{doelman2023laboratory}.}
   \end{figure}

Figure~\ref{fig:vvc_filtered_and_not} illustrates the impact of polarization leakage on VVC performance, showing that even small amounts of leakage can obscure planetary signals.
Classically, polarization leakage is mitigated using polarization filtering, which employs a sandwich of linear polarizers and quarter-wave plates around the VVC mask\cite{mawet2010VectorVortex}. However, this method significantly reduces throughput, blocking half of the incoming light. Additionally, the alignment of polarizers and wave plates must be precise, which can introduce further alignment errors and affect performance. Further, the wave plates need to be achromatic if the instrument is to work with large spectral bandwidths.

The SUPPPPRESS project aims to address polarization leakage more effectively than traditional polarization filtering methods. By developing multi-grating vector vortex coronagraphs (mgVVCs)\cite{doelman2020minimizing}, the project seeks to reduce leakage through diffraction grating techniques. This technique uses sets of polarization gratings to split and realign light based on its polarization state, significantly minimizing leakage without the substantial light loss associated with the traditional filtering method.

The multi-grating approach promises to suppress on-axis polarization leakage to levels as low as $10^{-10}$, making it a highly promising solution for future HCI missions. In the following section, we detail how polarization gratings can be used to achieve this.

\section{Multi-grating vector vortex coronagraphs}
\label{sec:mgVVCs}

The development of multi-grating vector vortex coronagraphs (mgVVCs) represents a significant advancement in addressing the challenges of polarization leakage inherent in traditional VVCs. By incorporating multiple diffraction gratings into the coronagraph design, these instruments can more effectively manage and reduce polarization leakage, thereby enhancing overall contrast performance. This section explores the principles of diffraction by polarization gratings, the implementation and benefits of double-grating VVCs (dgVVCs), and the further improvements achieved with triple-grating VVCs (tgVVCs), illustrating how each configuration contributes to the suppression of polarization leakage.

\subsection{Diffraction by a pair of polarization gratings}
\label{subsec:polarization-grating-pair}

A polarization grating (PG) is a special type of diffraction grating that manipulates light based on its polarization state. Identical to the VVC, a polarization grating introduces geometric phase but instead of a continuously rotating half-wave axis in the grating direction to create a vortex phase ramp, it introduces an opposite tilt for the two circular polarization states. Therefore, an ideal polarization grating with perfect half-wave retardance has only the $\pm 1$ orders and each order has one circular polarization state. The half-wave retardance also causes a flip in the polarization state of the light. Deviations from half-wave or patterning errors introduce a zero-order, and the latter also introduces higher-order diffraction. We note that unpolarized light has equal amounts of (incoherent) circularly polarized light, while linearly polarized light can be described as two coherent states of circularly polarized light. Therefore, both unpolarized and linearly polarized light produce the $\pm 1$ orders. To conclude, when light passes through a polarization grating, it is split into two beams with opposite polarization states, each diffracted at opposite angles. 

\begin{figure}
    \centering
   \includegraphics[width = 0.5\textwidth]{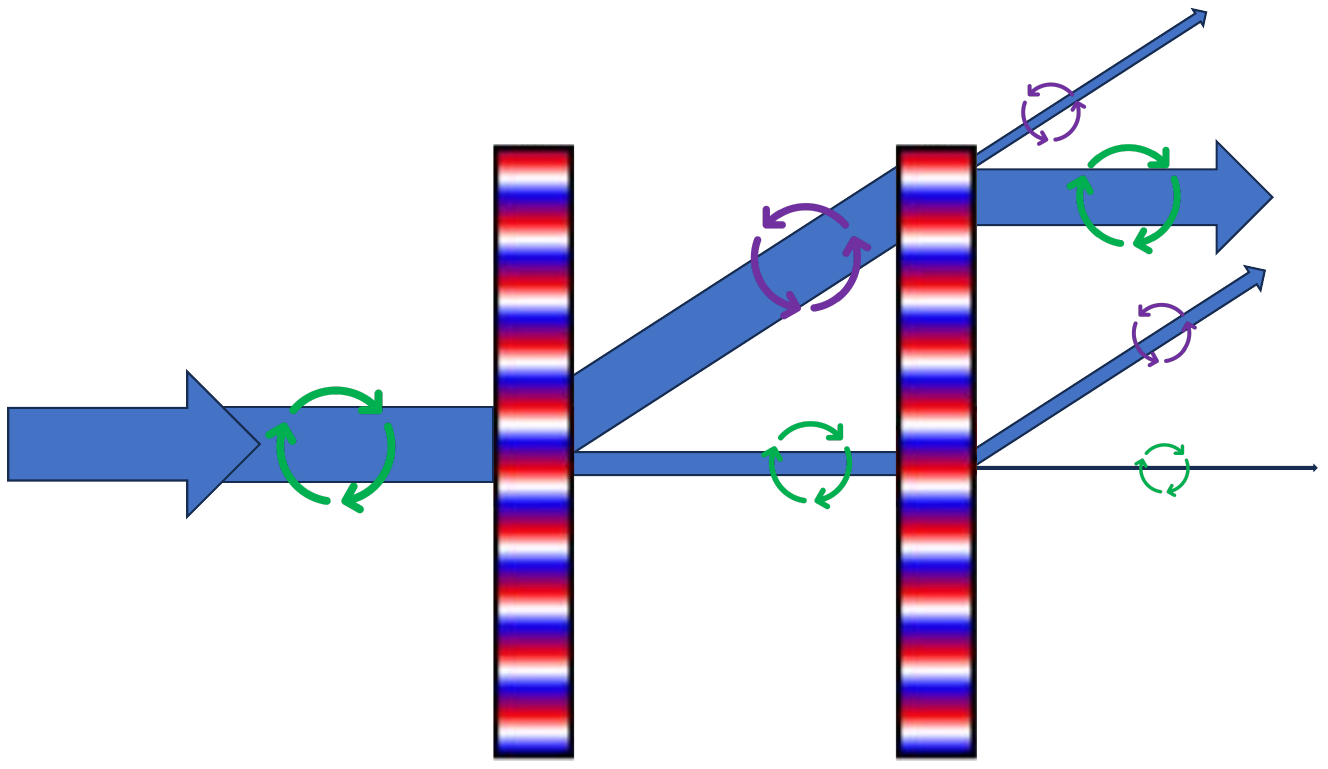}
   \caption[Figure] 
   {\label{fig:double_grating_half_beams} 
    Diffraction process of polarized light as it passes through a pair of polarization gratings. The thickness of the beams corresponds roughly to the amount of light in each beam. Initially, the incoming light is split based on its polarization state (indicated by circular arrow direction and color) by the first polarization grating. Here, we only show the resulting beams for one polarization state. The first grating diffracts the light, causing a flip in polarization. The diffracted light, now with altered polarization, proceeds to the second polarization grating. The second grating applies an opposite diffraction effect, which realigns the light back to its original axis. The leakage from the first polarization grating becomes the on-axis beam for the second grating, where it gets diffracted off-axis. There is some leakage associated with this as well, but it is significantly less than the original leakage through the first grating. There is also a leakage term associated with the main beam diffraction at the second grating, but that also goes off axis and ends up outside of the pupil. For an illustration of the effect on both polarization states, see Fig.~\ref{fig:dgVVC_combo}, right. Adapted from Doelman et al. (2020)\cite{doelman2020minimizing}.}
   \end{figure}

In a pair of polarization gratings, as illustrated for one polarization state in Fig.~\ref{fig:double_grating_half_beams}, the first grating diffracts the incoming polarized light and flips its polarization state. The diffracted light, now with an altered polarization, proceeds to the second polarization grating. The second grating applies an opposite diffraction effect due to the inverted polarization state, which realigns the light back to its original orientation and polarization state.
The first polarization grating produces a leakage term that continues on-axis without being properly diffracted. When this leakage light encounters the second polarization grating, now acting as the main beam on this second grating, it is primarily diffracted off-axis. This effectively removes the original leakage from the main beam path. The main beam, having been correctly diffracted by both gratings, remains on-axis. This dual-grating setup thus minimizes polarization leakage by ensuring that the majority of the leakage light is directed away from the main beam path.

\subsection{The double-grating VVC}

\begin{figure}
    \centering
   \includegraphics[width = \textwidth]{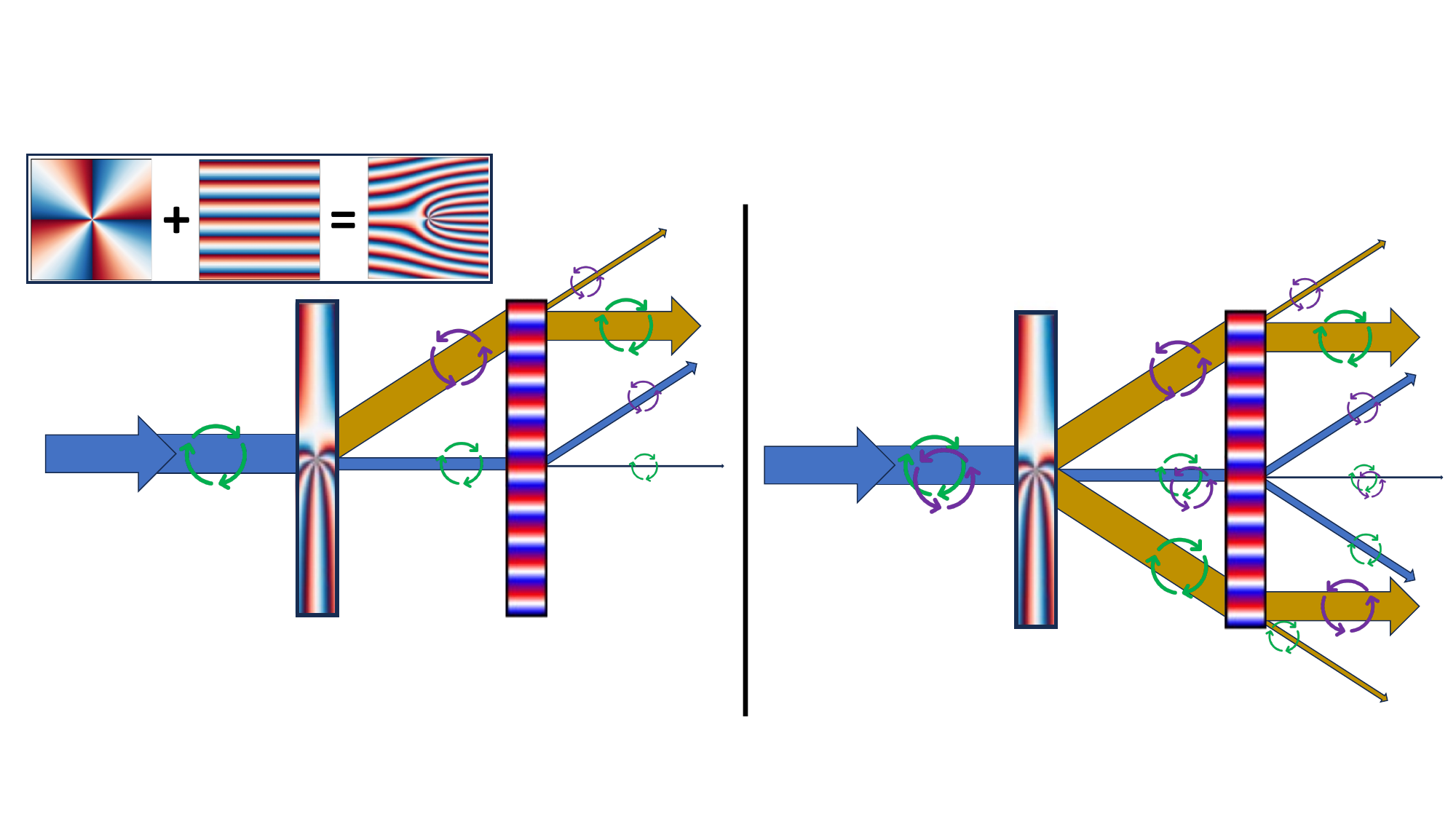}
   \caption[Figure] 
   {\label{fig:dgVVC_combo} 
    Integration of polarization gratings with vector vortex coronagraphs to form mgVVCs, here a dgVVC. \textit{Left:} The first polarization grating, combined with the phase pattern of a vector vortex coronagraph (sum shown in inlet on top left) diffracts the incoming light based on its polarization state. The light affected by the polarization grating also acquires the phase of the VVC, forming the coronagraphed beam realigned back to the original axis by the second grating (large orange beam). The thickness of the beams corresponds roughly to the amount of light in each beam. \textit{Right:} When the orthogonal polarization state is added (unpolarized light), the same diffraction process occurs with opposite sign. This dual-grating setup significantly reduces the polarization leakage term. Adapted from Doelman et al. (2020)\cite{doelman2020minimizing}.}
   \end{figure}

Building on the principles of diffraction by a pair of polarization gratings in the previous section, the double-grating VVC (dgVVC) is a VVC with the ability to significantly reduce polarization leakage. The first component of the dgVVC is a unique element that combines the phase pattern of a VVC with a polarization grating (see top left in Fig.~\ref{fig:dgVVC_combo}). This combination creates a complex pattern where the phase ramp of the VVC and the diffraction effect of the grating are superimposed.
When polarized light enters the dgVVC, it first encounters this combined phase pattern. The light is diffracted based on its polarization state, and its polarization is flipped by the grating component. Simultaneously, the same light acquires the phase of the VVC, which is crucial for nulling the starlight and allowing the planetary light to pass through. This light is marked in orange in Fig.~\ref{fig:dgVVC_combo}, left.
After passing through the first component, the light, now with an altered polarization state and the VVC phase, proceeds to the second polarization grating. This second grating realigns the main beam back to its original axis as described in Sec.~\ref{subsec:polarization-grating-pair}.

The primary advantage of this configuration is that the light that acquires the VVC phase is effectively managed through the dual-diffraction process, ensuring that it remains on-axis and retains the desired phase modulation. In contrast, the leakage term, which continues on-axis after the first diffraction, is redirected off-axis by the second grating, significantly reducing its impact on the system.

This dual-grating setup combined with the VVC, therefore, achieves a significant reduction in polarization leakage without the substantial light loss associated with traditional polarization filtering methods. By effectively managing the diffraction and alignment of the beams, the dgVVC enhances the contrast performance of the coronagraph, making it a powerful tool for HCI applications.

\subsection{The triple-grating VVC}

The dgVVC significantly reduces polarization leakage by using a pair of polarization gratings to diffract and realign the light, but the performance can be further enhanced. By adding an additional polarization grating, the triple-grating vector vortex coronagraph (tgVVC) offers an even more effective solution for managing and reducing leakage.

In the tgVVC design, the first component is similar to the dgVVC, where the combined phase pattern of the VVC and a polarization grating diffracts the incoming light, flipping its polarization state while imparting the necessary phase shift. But now, instead of stopping at two gratings, a third grating is introduced. The grating frequencies between the full set of gratings is adjusted to make sure the light that is affected by all three components is diffracted back on axis. This additional grating also performs another round of diffraction on the light originally leaking through the first component, further directing any remaining leakage off-axis.

\begin{figure}
    \centering
   \includegraphics[width = \textwidth]{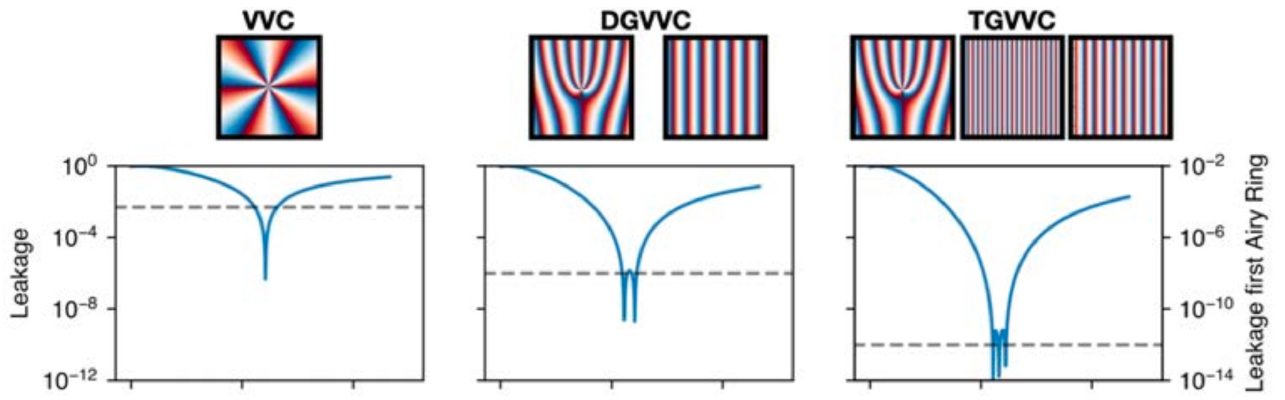}
   \caption[Figure] 
   {\label{fig:leakage_comparison} 
    Comparison of the leakage terms and their intensities for different VVC configurations. The left plot shows the leakage term for a standard VVC, which is around $10^{-4}$. The center plot illustrates the leakage term using a dgVVC, achieving levels between $10^{-6}$ and $10^{-7}$. The right plot demonstrates the leakage in a tgVVC, at below $10^{-10}$.}
   \end{figure}

Figure~\ref{fig:leakage_comparison} illustrates the effectiveness of this approach by comparing the leakage terms for a standard VVC, a dgVVC, and a tgVVC. The left plot shows the leakage term for a standard VVC, which is around $10^{-4}$. The center plot illustrates the leakage term using a dgVVC, achieving levels between $10^{-6}$ and $10^{-7}$. The right plot demonstrates the leakage in a tgVVC, achieving a normalized intensity level below $10^{-10}$. This remarkable reduction showcases the superior performance of the tgVVC design in suppressing polarization leakage.

By implementing a triple-grating system, the tgVVC achieves unparalleled leakage suppression, making it an optimal choice for HCI applications, especially for missions requiring extremely high contrast levels, such as detecting Earth-like exoplanets.

\section{The SUPPPPRESS project}
\label{sec:suppppress}

The SUPPPPRESS (Substantiating Unique Patterned Polarization-sensitive Polymer Photonics for Research of Exoplanets with Space-based Systems) project aims to address the significant challenge of polarization leakage in VVCs through the development and implementation of mgVVCs. Utilizing advanced liquid-crystal technology, the project seeks to enhance the performance of coronagraphs to meet the stringent requirements of HCI missions such as HWO. This project is a collaborative effort led by Leiden University, in a consortium with academic and industry partners SRON, cosine, and LESIA at Paris Observatory.

\subsection{Scope and goals}

The primary scope of the SUPPPPRESS project involves the detailed design, manufacturing, testing, and validation of mgVVCs using liquid-crystal technology. The project focuses on creating comprehensive designs for liquid-crystal masks optimized for HCI missions like HWO. This includes refining phase patterns and alignment techniques to minimize polarization leakage and maximize throughput.

Precision manufacturing of liquid-crystal components is crucial, and the project collaborates with Colorlink Japan\footnote{\url{https://www.colorlink.co.jp/en/}} to produce high-quality liquid-crystal masks. These components undergo rigorous testing to ensure their performance in terms of retardance accuracy, uniformity, and durability. Component-level tests that will be performed in the Netherlands include polarization microscope assessments to measure the retardance and uniformity of the liquid-crystal layers and far-field diffraction analysis.

The assembled mgVVC coronagraph will undergo HCI tests using advanced testbeds, such as the THD2 testbed at Paris Observatory. These tests evaluate the coronagraph's performance under realistic conditions, focusing on contrast levels and response to various light sources. One of the project’s goals is to achieve a contrast level of $10^{-9}$ in laboratory settings within the next two years, a benchmark essential for detecting and characterizing Earth-like exoplanets.

Additionally, the project includes a series of environmental tests to ensure the mgVVC coronagraph can operate effectively in the harsh conditions of space. These tests will simulate the thermal, vacuum, and radiation environments of space to verify the durability and reliability of the liquid-crystal components and the overall coronagraph assembly.

\subsection{Previously manufactured patterns and previous results}

The SUPPPPRESS project builds on the foundation of previously manufactured VVC patterns, utilizing liquid-crystal technology to achieve high precision and performance. Initial prototypes, including the first tgVVCs, were manufactured by ImagineOptix\cite{doelman2023laboratory}. The manufacturing process involved direct-write techniques to create precise liquid-crystal patterns on substrates. The tgVVC prototype consisted of multiple layers, including a forked grating and various grating pitches arranged in a 1:3:2 ratio to keep the main, coronagraphed beam on axis. 

\begin{figure}
    \centering
   \includegraphics[width = 0.8\textwidth]{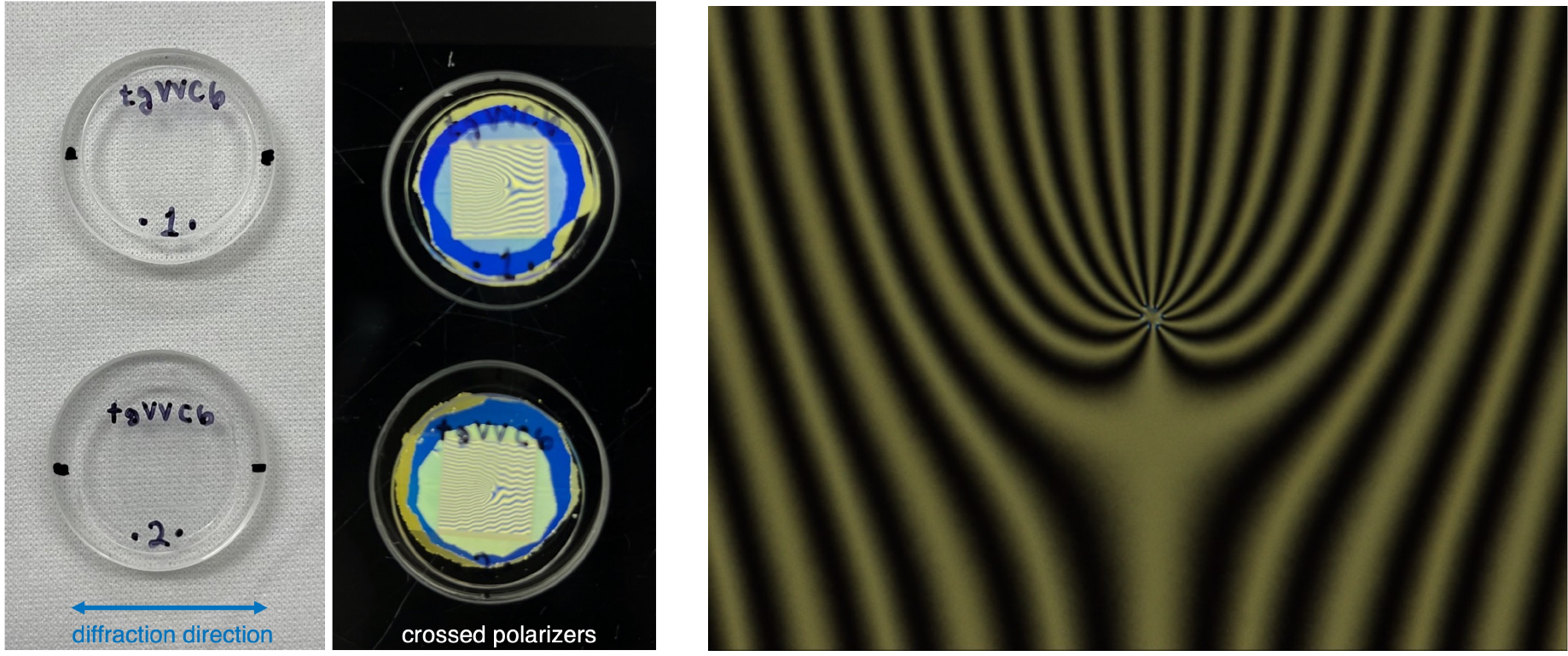}
   \caption[Figure] 
   {\label{fig:first_manufactured_patterns} 
    \textit{Left:} First prototype of the tgVVC as manufactured by ImagineOptix. The pattern inaccuracies, retardance errors, and alignment errors result in a non-vortex pattern between crossed polarizers. Adapted from Doelman et al. (2023)\cite{doelman2023laboratory}. \textit{Right:} One of the first liquid-crystal mgVVC patterns manufactured by Colorlink Japan.}
   \end{figure}

Figure~\ref{fig:first_manufactured_patterns} (left) shows the first prototype of the tgVVC, highlighting the intricate patterning and alignment required to achieve the desired optical properties.
Pattern accuracy and retardance uniformity are critical for the effectiveness of tgVVCs. The initial prototypes showed some deviations from the ideal half-wave retardance, which contributed to polarization leakage. Retardance errors of up to 5\% were observed in some areas of the masks, leading to non-ideal phase shifts and subsequent leakage. Additionally, the alignment of the grating patterns with the VVC phase masks was found to be within 2-3 microns accuracy, which, while precise, still left room for improvement. These inaccuracies highlighted the importance of refining the manufacturing process to ensure tighter control over the liquid-crystal patterning and alignment.

\begin{figure}
    \centering
   \includegraphics[width = \textwidth]{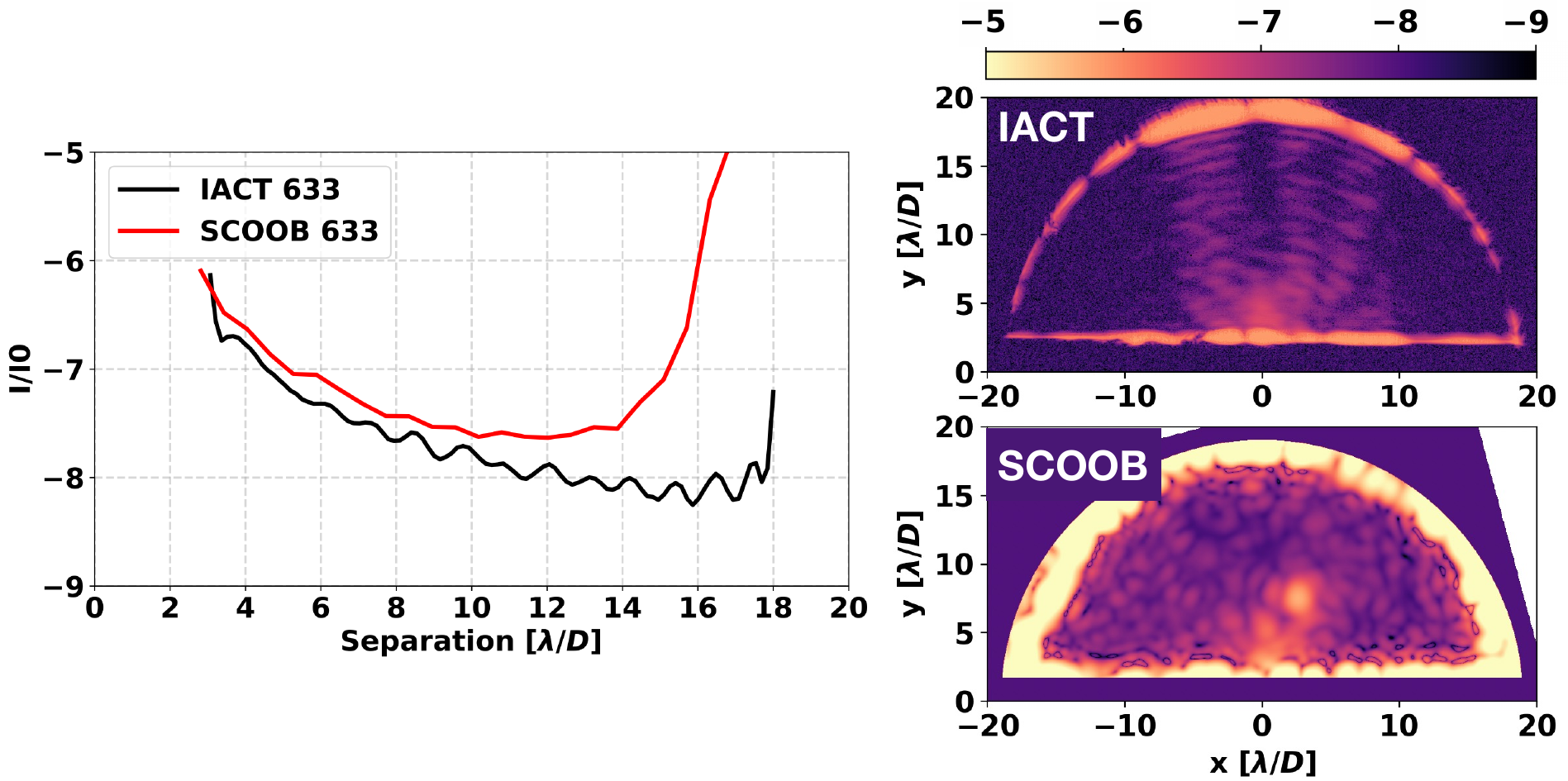}
   \caption[Figure] 
   {\label{fig:previous_lab_results} 
    Pre-SUPPPPRESS testbed results published by Doelman et al. (2023)\cite{doelman2023laboratory}. The same tgVVC mask was tested both in the in-air testbed at JPL (IACT) and the SCoOB testbed at the University of Arizona. \textit{Left:} Normalized intensity of the experiments on either testbed. \textit{Right:} Best dark holes after closed-loop contrast correction. While the overall results are comparable, notable differences in the dark hole morphology are evident. The JPL testbed exhibits striping patterns, whereas the SCoOB testbed shows a more speckled appearance. These differences are attributed to the use of different control techniques: model-based Electric Field Conjugation (EFC) at JPL and implicit EFC at SCoOB.}
   \end{figure}

The performance of these prototypes was tested in various settings, including the In-Air Coronagraphic Testbed\cite{Baxter2021DesignAndComm} (IACT) at NASA’s Jet Propulsion Laboratory and the Space Coronagraph Optical Bench\cite{Ashcraft2022TheSpaceCoro,vanGorkom2022TheSpaceCoro} (SCoOB) at the University of Arizona. For instance, tests on the IACT revealed an average contrast of $2 \times 10^{-8}$ between 3--18 $\lambda/D$, while the SCoOB testbed achieved a contrast of $6 \times 10^{-8}$ between 3--14 $\lambda/D$. These results, shown in Fig.~\ref{fig:previous_lab_results}, indicate the tgVVC's potential for HCI, although they also highlight the need for further refinements to reach the desired $10^{-10}$ contrast levels necessary for detecting Earth-like exoplanets.
Despite the promising results, the tests also identified several challenges and areas for improvement. The IACT testbed exhibited structured dark-hole regions, likely due to model inaccuracies in the Electric Field Conjugation (EFC) algorithm, suggesting that the tgVVC prototypes might have minor manufacturing errors. On the other hand, the SCoOB testbed, using implicit EFC (iEFC), showed a more uniform speckle field, albeit with a lower overall contrast. These discrepancies underscore the importance of refining both the manufacturing processes and the control algorithms to fully assess the potential of tgVVCs.

Overall, the previously manufactured patterns and test results provide a solid foundation for the SUPPPPRESS project. By addressing the identified issues and leveraging advanced manufacturing techniques, the project aims to produce tgVVCs capable of achieving the stringent performance requirements for future HCI missions.

\subsection{Project timeline}

\begin{figure}
    \centering
   \includegraphics[width = \textwidth]{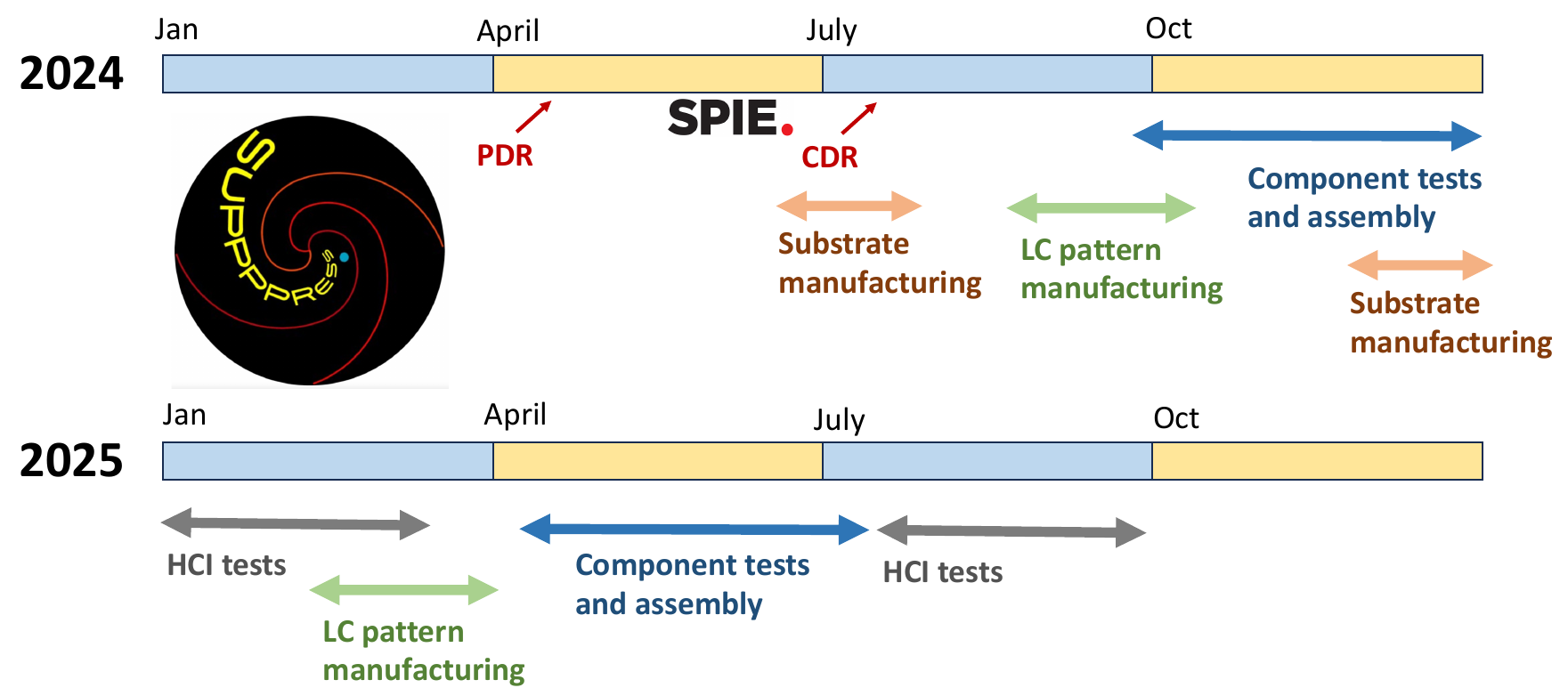}
   \caption[Figure] 
   {\label{fig:timeline} 
    Timeline of the SUPPPPRESS project. At the time of this conference, we have passed PDR and ordered a first batch of substrates.The first patterns will be manufactured this fall and will be tested before the end of the year. Component level testing will preceed tests on a HCI testbed. From the component-level tests with polarization microscopes, and through the assembly and the gluing, we anticipate valuable lessons learned to fold into a second string of manufactured patterns.}
   \end{figure}

The total duration of the SUPPPPRESS project is two years and started in October 2023, with an overview of its timeline shown in Fig.~\ref{fig:timeline}. We passed the preliminary design review (PDR) in April 2024, two months prior to this SPIE conference. We will go through critical design review (CDR) in July 2024. A first substrate order has been put in, which will be used for the manufacturing of the first set of liquid-crystal patterns in Japan in the fall of 2024. Between October and December 2024, component-level testing on these patterns will be performed before they are assembled into the full liquid-crystal mask assembly (LCMA), the full mgVVC. Since we expect to learn more about the patterning accuracy and retardance errors of the LCMAs whose acquisition and manufacturing is currently in process, we aim to launch a new set of manufactured substrates and liquid-crystal patterns by the end of this year, capitalizing on the lessons learned from the first batch. The HCI tests of the first batch will be started on the THD2 testbed in Paris in the beginning of 2024, as described in the following section.

\section{The THD2 testbed as a European prototyping facility}
\label{sec:THD2-facility-Europe}

The THD2 testbed at the Paris Observatory is a state-of-the-art facility designed for HCI research and development. As a critical part of the SUPPPPRESS project, the THD2 testbed provides a stable and well-characterized environment for testing and validating advanced coronagraph designs like the mgVVCs. This section details the capabilities of the THD2 testbed and its role in supporting the SUPPPPRESS project's goals.

\subsection{The THD2 testbed for SUPPPPRESS}

The THD2 testbed in Paris has a long heritage, having been operational for more than 10 years\cite{Baudoz2018OptimizationPerformanceMultideformable}. This extensive use has allowed researchers to thoroughly understand its characteristics, such as micro vibrations, stability, ghosts, and polarization aberrations. The well-known behavior of the THD2 testbed in these areas is a significant advantage for the SUPPPPRESS project, particularly in understanding and improving polarization-based vortex coronagraphs. Recently, measurements of polarization aberrations have been performed on this testbed\cite{Baudoz2024Polarization}, which are directly relevant to the SUPPPPRESS project. These measurements help in refining the design and testing of the mgVVCs to achieve better performance in terms of polarization leakage.

\begin{figure}
    \centering
   \includegraphics[width = \textwidth]{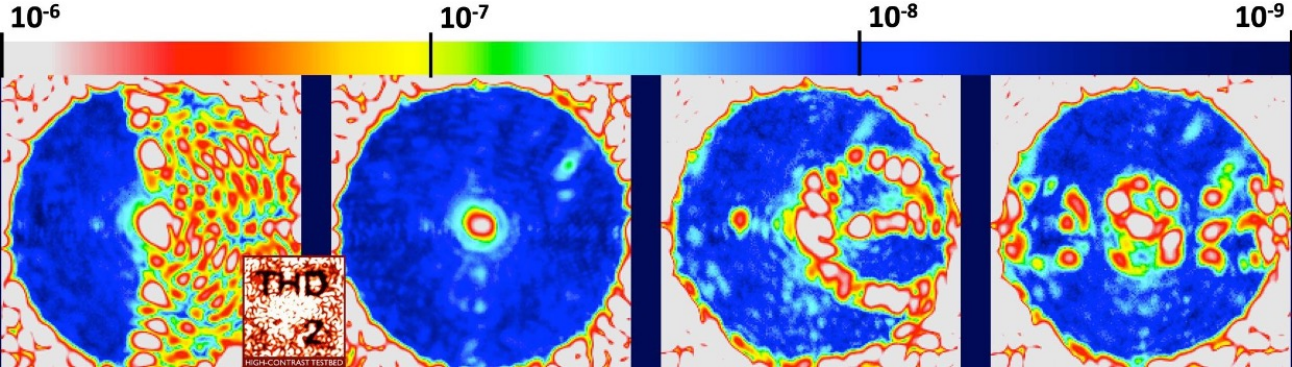}
   \caption[Figure] 
   {\label{fig:thd2-DHs} 
    Examples of DHs on THD2 using an FQPM. The THD2 testbed can reach $4-5 \times 10^{-9}$ contrast in air. It capitalizes on different control techniques that have been tested over the years, including pair-wise probing, electric field conjugation and the self-coherent camera.\cite{Potier2020_THD2}}
   \end{figure}

The THD2 testbed is equipped with two deformable mirrors and a tip-tilt mirror for active wavefront sensing and control, enabling high precision in manipulating the wavefront. In terms of performance, the THD2 testbed can reach contrasts of $4-5 \times 10^{-9}$ in air, a critical feature for testing HCI instruments. The testbed supports various control techniques developed and refined over the years, including EFC with classic pairwise probing and the self-coherent camera, a form of iEFC\cite{Potier2020_THD2,Herscovici-Schiller2018ExperimentalValidationNonlinear,Mazoyer2014SCC}. Some DH examples on the THD2 testbed are shown in Fig.~\ref{fig:thd2-DHs}.

The THD2 testbed has been used to test a large number of coronagraphs (see Tab.~\ref{tab:coronagraphs-on-thd2}), leveraging collaborations with international partners. This includes a variety of phase-mask technologies, such as the four quadrant phase mask\cite{Bonafous2016Development}, vector vortex masks provided by different manufacturers, and the wrapped vortex phase mask\cite{Galicher2020AFamily}, a scalar vortex mask that aims to improve achromaticity. The extensive history of coronagraph testing on THD2 allows for a robust ``apples to apples’’ comparison among different coronagraph types.
\begin{table}[ht]
    \centering
    \caption{Coronagraphs tested and used on THD2 currently and in the past.}
    \label{tab:coronagraphs-on-thd2}
    \begin{tabular}{lll}
        \toprule
        \textbf{Coronagraphic Components} & \textbf{Years} & \textbf{Collaboration} \\
        \midrule
        Four Quadrant Phase Mask & 2006 $\rightarrow$ & GEPI, France \\
        Multi-FQPM & 2009-2016 & GEPI, France \\
        Apodized Dual Zone Phase Mask & 2012-2016 & LAM, France \\
        Six Level Phase Mask & 2015-2018 & GEPI, France - Univ. of Shanghai, China \\
        Vector Vortex Phase Mask (photonic crystal) & 2015-2017 & NAOJ, Japan \\
        Vector Vortex Phase Mask (liquid-crystal polymer) & 2017 $\rightarrow$ & LESIA, France, Leiden Univ., Netherlands \\
        Eight Octant Phase Mask (photonic crystal) & 2015-2017 & Univ. of Hokkaido, Japan \\
        Achromatic Wrapped Phase Mask & 2016 $\rightarrow$ & LESIA, France \\
        Phase-Apodized-Pupil Lyot Coronagraph (PAPLC) & 2016-2020 & Leiden Univ., Netherlands \\
        Roman Space Telescope Coronagraphs & 2019-2021 & LAM, France \\
        Multiple grating Vector Vortex Coronagraph & 2023-2025 & Leiden Univ., SRON, Netherlands \\
        \bottomrule
    \end{tabular}
\end{table}
Within the SUPPPPRESS project, the goal is to compare the performance of the new mgVVCs against these established types to comprehensively characterize their capabilities. This comparison will be critical in validating their effectiveness and understanding how they can be optimized for future HCI missions.

\subsection{THD2 Testbed upgrade}

The THD2 testbed is currently undergoing significant upgrades to enhance its capabilities as a prototyping facility for the international HCI community, especially within Europe. These upgrades aim to make the testbed more versatile and capable of supporting a broader range of research activities.

A series of hardware upgrades are being implemented to improve the testbed's functionality. This includes the installation of a new supercontinuum light source, which will serve as a broadband light source and is expected to arrive in the fall of 2024. Additionally, a new camera is about to be delivered and installed, further enhancing the testbed’s imaging capabilities. A new tip-tilt mirror has been upgraded to replace the old one in April of 2024, which had reached the end of its operational life. Furthermore, a new PC was installed in the fall of 2023 to support the enhanced control infrastructure.

A major part of the upgrade focuses on the control infrastructure. The current control systems are being replaced with a C++/Python implementation, developed as part of the catkit2 library\cite{Por2024catkit2}. This upgrade is being carried out in collaboration with the HiCAT team at the Space Telescope Science Institute\cite{Soummer2024HiCAT}. This new control infrastructure aims to improve the efficiency and flexibility of the testbed, allowing for more precise and advanced control of the experimental setups. The goal is to create a platform that allows for the flexible implementation of new algorithms in a high-level language or graphical user interface (GUI). This includes integrating machine learning-based control strategies and exploring new methods for broadband wavefront sensing and control\cite{Gutierrez2024}.

The overall objective of these upgrades is to establish the THD2 testbed as a versatile and accessible HCI prototyping facility. By integrating advanced hardware and software solutions, the testbed will support a wide range of research activities and collaborations, driving forward advancements in coronagraph technology and HCI.

\subsection{The European high-contrast imaging community}

The SUPPPPRESS project is part of a larger European initiative to advance HCI technology. In March of 2024, a two-day workshop was organized in Paris to unite the European HCI community\footnote{\url{https://hcieurope.sciencesconf.org/}}. Experts from European institutions were brought together to discuss ongoing research and future directions. Significant past and ongoing contributions were recognized, but the need for better coordination within the European context was highlighted, especially compared to efforts abroad related to HWO. The workshop was attended by two ESA representatives to the Science, Technology and Architecture Review Team (START)\footnote{\url{https://www.greatobservatories.org/hwo-start}}, facilitating vital connections to activities in relation to HWO. One key outcome was the decision to continue these discussions and collaborations with another workshop planned for next year, around March 2025.

To improve communication and collaboration within the European instrumental HCI community, a new Slack workspace was established, providing a centralized communication platform for European researchers. All interested colleagues with affiliations at European institutions are encouraged to join and participate actively. This initiative aims to create an inclusive, well-coordinated space for European HCI research, leveraging collective expertise to drive forward shared goals.

\section{Summary, conclusions and outlook}
\label{sec:summary-conclusions-outlook}

The SUPPPPRESS project addresses the critical issue of polarization leakage in standard vector vortex coronagraphs (VVCs), which limits their effectiveness in HCI for exo-Earth detection. By combining VVC patterns with multiple polarization gratings, the project aims to develop tgVVCs to significantly reduce the leakage.

Simulated results indicate that standard VVCs exhibit a leakage term around $10^{-4}$, while dgVVCs can reduce this to $10^{-6}$ to $10^{-7}$, and tgVVCs can achieve levels below $10^{-10}$. These simulations show the potential of multi-grating configurations to substantially reduce leakage, thus providing a promising solution for space-based coronagraphy that also enables observations with polarimetry since polarization filtering is not required anymore.

Reducing polarization leakage in this way has significant scientific implications. Achieving higher contrast levels enables the direct imaging and characterization of Earth-like exoplanets, which are crucial for understanding planetary systems and the potential for life beyond our solar system. By effectively managing the polarization leakage, mgVVCs will allow for more detailed studies of exoplanet atmospheres, compositions, and potentially even surface conditions. As highlighted by Vaughan et al. (2023)\cite{vaughan2023chasing}, polarimetry is necessary for detecting scattering phenomena such as ocean glint and rainbows. These optical features can reveal critical information about planetary habitability, including surface liquid water and atmospheric conditions.

The SUPPPPRESS project, funded by ESA, is focused on refining these coronagraph designs, with a goal of achieving an initial laboratory contrast level of $10^{-9}$. The project includes detailed design, manufacturing, and validation of liquid-crystal components, which are then integrated and tested on the THD2 testbed at Paris Observatory. This testbed, equipped with advanced hardware and upgraded control systems, provides a stable environment to validate the performance of mgVVCs.

Until the end of this year, SUPPPPRESS activities will focus on manufacturing the first set of liquid-crystal patterns and conducting component-level testing. This includes precise polarization microscope assessments, far-field diffraction analysis and alignment tests. Following these tests, the liquid-crystal components will be assembled into the full multi-grating VVC configuration for HCI and environmental testing.

The THD2 testbed plays a crucial role in these efforts, offering a controlled environment to test the coronagraph designs and improve their performance. This testbed's well-characterized stability and advanced control techniques are essential for achieving the project's ambitious goals.

\acknowledgments 
This work was supported by the European Space Agency (ESA) under the tender number \linebreak TDE-TEC-MOO AO/1-11613/23/NL/AR.

\bibliography{references}
\bibliographystyle{spiebib}

\end{document}